\documentclass[a4paper]{jpconf}
\usepackage{graphicx}
\begin{document}
\title{A modular mini-pad photon detector prototype for RICH application at the Electron Ion Collider}

\author[add-trieste]{J.~Agarwala$^{4}$, C.D.R.~Azevedo$^{2}$, C.~Chatterjee$^{3}$, A.~Cicuttin$^{4}$, P.~Ciliberti$^{3}$, M.L.~Crespo$^{4}$, S.~Dalla~Torre$^{1}$, S.~Dasgupta$^{1}$, M.~Gregori$^{1}$, 
S.~Levorato$^{1}$, G.~Menon$^{1}$, F.~Tessarotto$^{1}$, Y.X.~Zhao$^{1}$}

\address{$^1$INFN Trieste, Trieste, Italy} 
\address{$^2$University of Aveiro, Aveiro, Portugal} 
\address{$^3$University of Trieste and INFN Trieste, Trieste, Italy}
\address{$^4$ Abdus Salam ICTP, Trieste, Italy and INFN Trieste, Trieste, Italy }

\ead{shuddha.dasgupta@ts.infn.it}

\begin{abstract}
Experiments at the future Electron Ion Collider require excellent hadron identification in a broad momentum range, in harsh conditions. A RICH capable to fulfill the PID requirements of the EIC could use MPGD-based photon detectors with solid photocathodes for covering large surfaces at affordable cost, providing good efficiency, high resolution and compatibility with magnetic field.
Photon detectors realized by coupling THGEMs and Micromegas have been successfully operated at the RICH-1 detector of the COMPASS Experiment at CERN since 2016. A similar technology could be envisaged for an EIC RICH, provided a large improvement in the photon position resolution is achieved. 
An R\&D effort in this direction is ongoing at INFN Trieste. Few prototypes with smaller pixel size (down to 3 mm x 3 mm) have been built and tested in the laboratory with X-Ray and UV LED light sources. A modular mini-pad detector prototype has also been tested at the CERN SPS H4 beamline. 
New data acquisition and analysis software called Raven DAQ and Raven Decoder have been developed and used with the APV-25 based Scalable Readout System (SRS), for the modular mini-pad prototype tests.
\par
The main characteristics of the new mini-pad hybrid MPGD-based detector of single photons are described and preliminary results of  laboratory and beam tests are presented.

\end{abstract}


\section{Introduction}
\par

The Electron-Ion Collider (EIC) \cite{EIC}, will be the ideal accelerator facility to explore Quantum Chromo Dynamics (QCD) because of its unprecedented luminosity, energy range and beams of polarized electrons colliding with beams of either polarized nucleons or nuclei. Experiments at the EIC will have demanding requirements on Particle Identification (PID), in particular on hadron PID at high momenta, where the natural choice is the gaseous RICH technique, which usually needs long radiators. The challenge posed by the limited length of the radiator compatible with the collider experiments imposes to design a Ring Imaging Cerenkov Counter (RICH) with smaller focal length, therefore an improved spatial resolution of the Photon Detectors (PDs) is needed to preserve the resolution in the measurement of the Cherenkov angle.
\par
In the recent upgrade of the COMPASS
RICH-1 detector \cite{upgradeHybrid, upgradehybridnew} four MPGD based PDs, covering a total area of 1.4 $m^{2}$, were successfully
implemented and operated.
The architecture of the new COMPASS PDs consists of three gas multiplication stages: two THGEM layers
(the first of which is coated with CsI and acts as a reflective photocathode) and a Micromegas (MM);
the anode is segmented in square pads of 8 mm pitch and the signals are read via capacitive coupled pads embedded in
the anode PCB, by an APV-25 based front-end.


\section{The modular minipad prototype PD}

In view of fulfilling the requirements imposed to the future RICH at the EIC experiments 
a prototype similar to the COMPASS PDs has been designed and built: it has an active area of 10$\times$10 cm$^{2}$,
an anode segmented in 1024 square 3$\times$3~mm$^{2}$ pads having 3.5 mm pitch. 
The prototype is fully
modular, with front-end electronics and all services contained 
in the 10$\times$10~cm$^{2}$ active area:
detectors covering larger areas could be designed by multiple replica of the basic module represented by the prototype.

\begin{figure}
	\centering
	\includegraphics[width=0.9\linewidth]{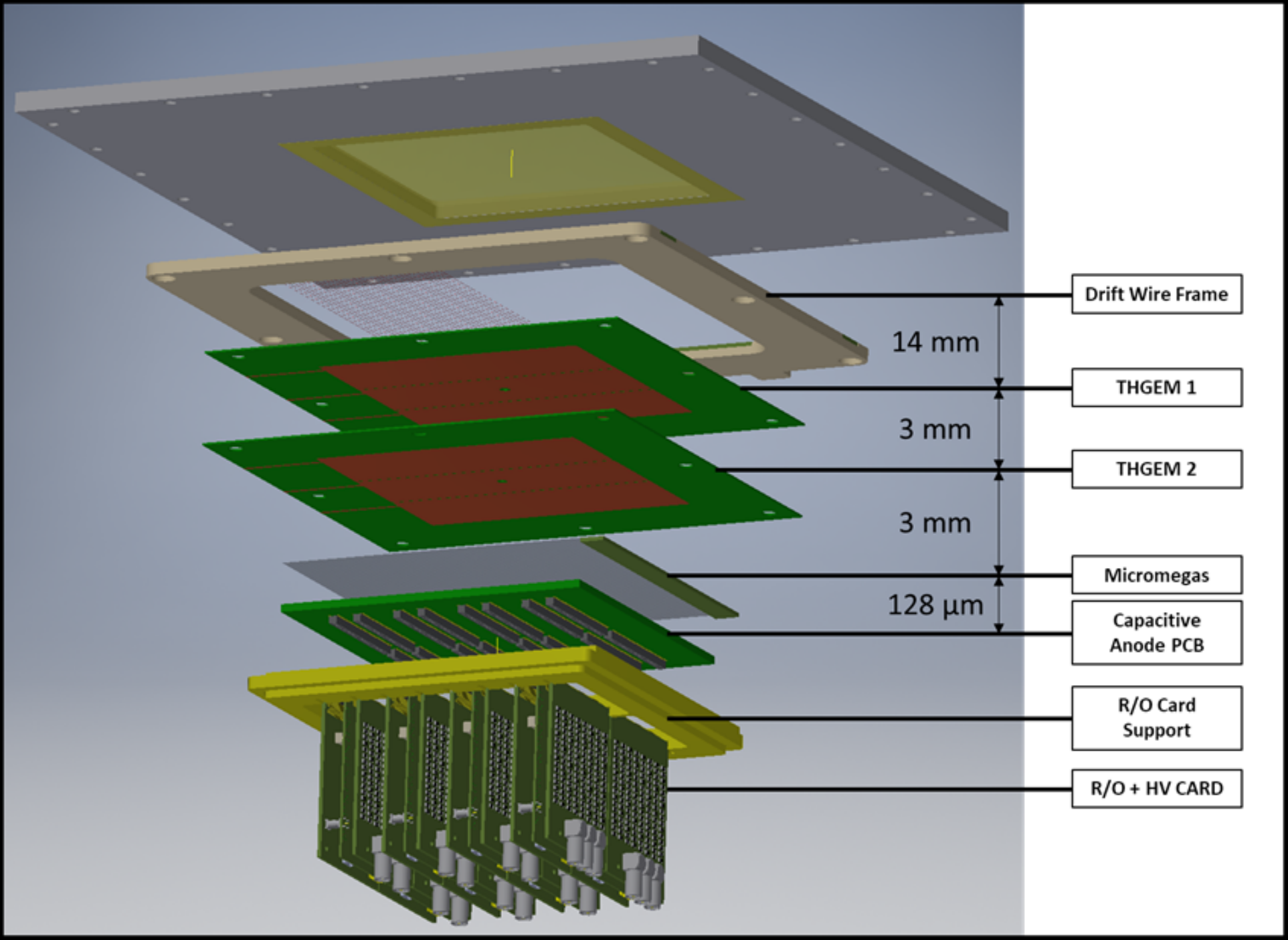}
	\caption{Schematic of the mini-pad prototype.}
	\label{fig:hybridminipadschematic}
\end{figure}

The internal structure of the modular minipad prototype is presented in Fig.\ref{fig:hybridminipadschematic} and consists of four layers:

\begin{itemize}
	\item a drift plane made of 100 $\mu$m diameter wires (CuBe Alloy 25, Au coated);
	\item a first THGEM (0.4~mm thick, 0.4`mm diameter holes with 0.8~mm pitch and no rim) 
	with CsI coating for UV photons conversion
	\item a second THGEM with the same geometrical parameters, mounted with holes staggered to provide maximal misalignment
	with respect to the holes of the first one;
	\item a MM built on a pad segmented anode with high granularity.
\end{itemize}

\subsection{THGEMs}

\par
\begin{figure}
	\centering
	\includegraphics[width=0.7\linewidth]{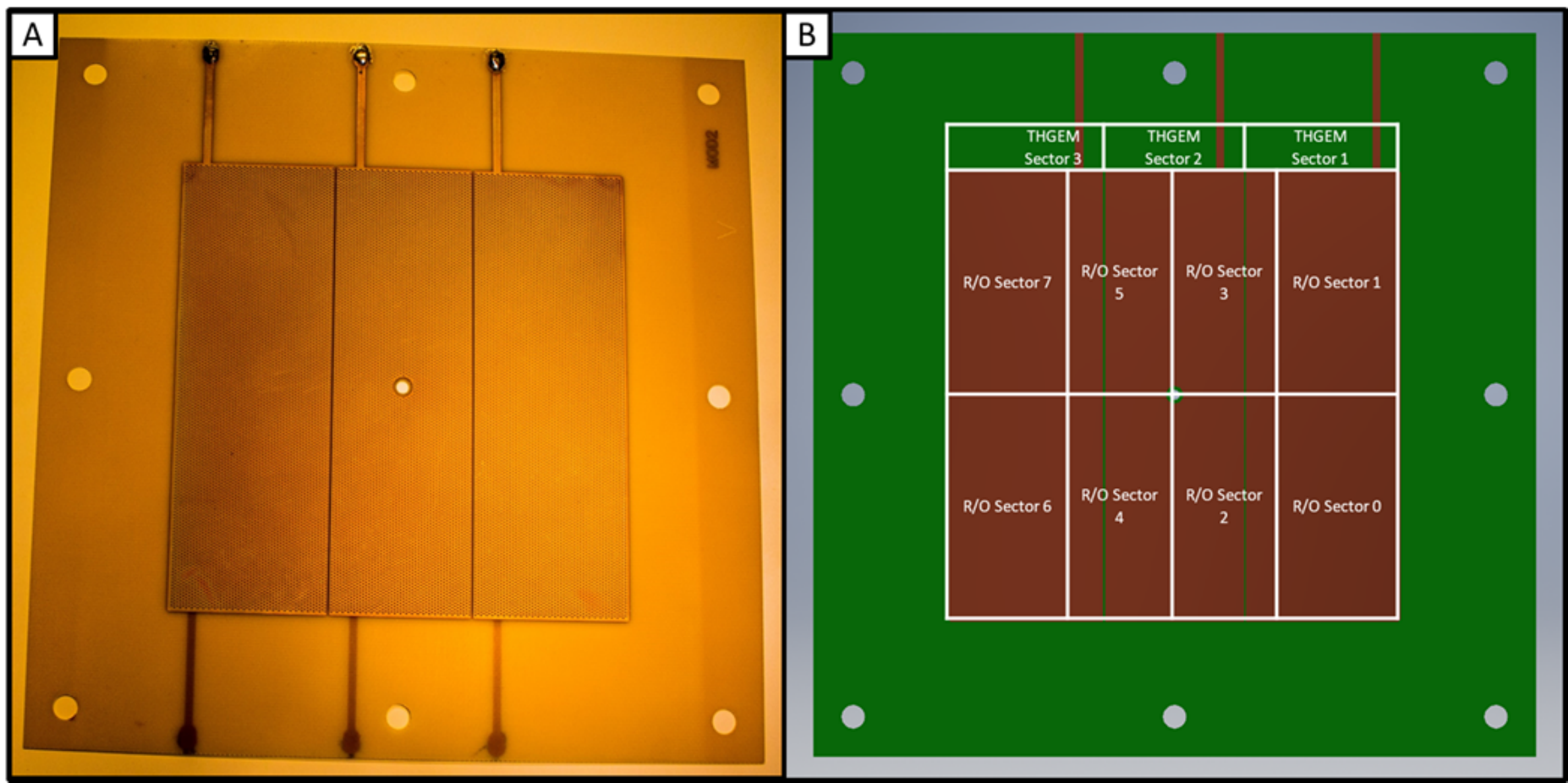}
	\caption{A. A THGEM PCB after the full treatment; B. Superposition of the schematic of readout sectors over the CAD schematic of a THGEM PCB.}
	\label{fig:thgemschematic}
\end{figure}

THGEMs \cite{THGEM_others} are standard Printed Circuit Boards (PCBs) with mechanically drilled patterned holes, each of which acts as a gaseous electron multiplier.
A seven-year R\&D at INFN Trieste has made them adequate for  
RICH applications \cite{THGEM_rd}. Eight THGEMs were produced for this prototype starting from raw material PCBs previously selected
for the COMPASS RICH-1 upgrade on the basis of their good thickness uniformity;
the top and bottom electrodes are segmented in three sectors  of 10$\times$3.1~cm$^{2}$ each
(Fig.~\ref{fig:thgemschematic}-A)

\par
The THGEMs were produced at ELTOS SpA and underwent the established post-production
treatment~\cite{Polishing} at the INFN Trieste Laboratory. The treatment consists in 
polishing with fine grain pumice powder, cleaning with high pressure water,
ultrasonic bath with a solution of pH 11 and drying in oven at
180$^\circ$C for 24h. 

\par
The THGEMs were characterized using a test chamber with an anodic plane segmented in 8
readout sectors of 28 $\times$ 56 mm${^2}$ (Fig.\ref{fig:thgemschematic}-B).
Ar:CO$_{2}$ 70:30 gas mixture and two X-ray sources ($^{55}Fe$ and AMPTEK Mini-X with
Au target and Cu filter) were used. After electrical strength tests, 
the gain variation under
continuous illumination at the nominal voltage was studied and
automated amplitude spectra were collected at different THGEM high voltage bias values;
discharge counting was performed: a discharge rate $<$ 10$^{-3}$ Hz at an effective
gain of 100 was observed for all THGEMs.

\par
\begin{figure}
	\centering
	\includegraphics[width=0.7\linewidth]{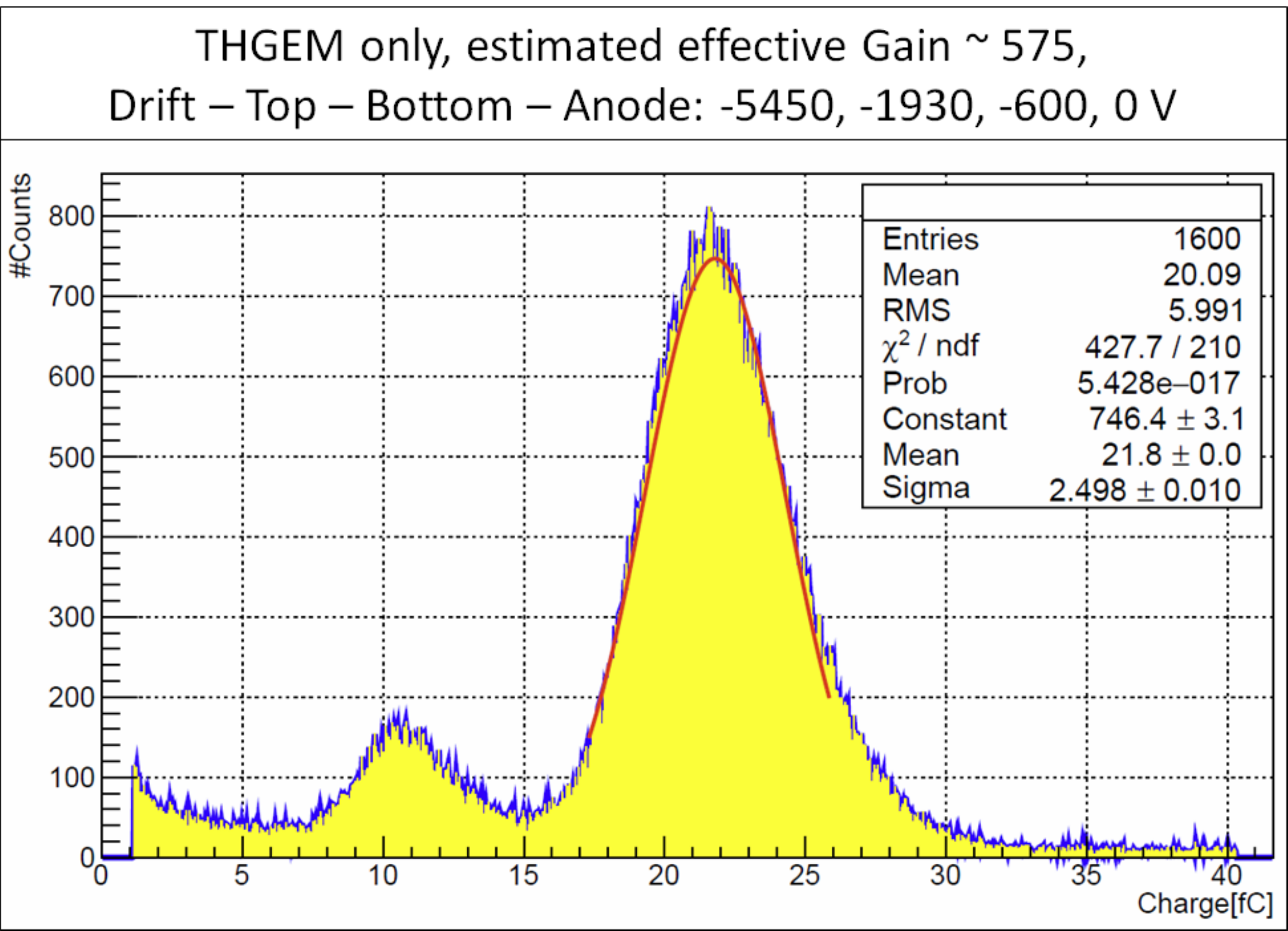}
	\caption{$^{55}$Fe Spectra of a single THGEM multiplication layer during characterization.}
	\label{fig:thgem55fespectra}
\end{figure}

The gain uniformity has been studied using $^{55}Fe$ X-ray amplitude spectra (Fig.\ref*{fig:thgem55fespectra}). Thanks to the pre-selection of the PCB material, an effective gain uniformity of $\sim{5}$\% was obtained. All eight produced THGEMs were validated. A Ni - Au coating to prepare four THGEMs for CsI layer deposition was performed.

\subsection{The minipad Micromegas prototype}

\par
\begin{figure}
	\centering
	\includegraphics[width=0.9\linewidth]{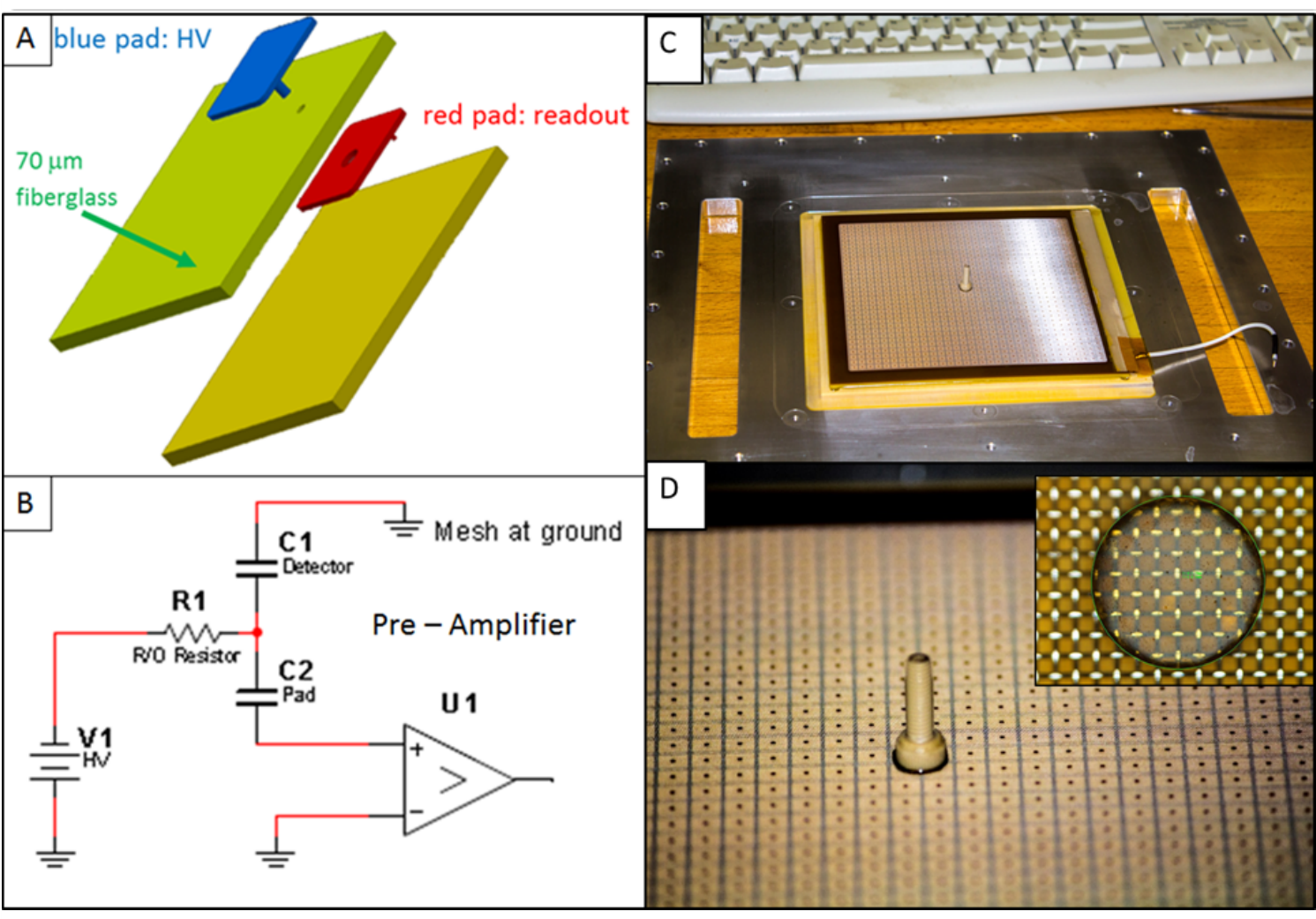}
	\caption{A: Exploded view of one single readout pad structure. B: The schematic of the circuit diagram of the capacitive anode principle. C: The MM side of the anode PCB. D: Zoom of the pad plane where the pillars are seen in the center of each pad. Inset: 
	Picture by microscope 
	of a pillar.}
	\label{fig:micromegasscheme}
\end{figure}

Two MMs (Fig.\ref{fig:micromegasscheme}) were produced at CERN using standard bulk technology \cite{MM}. They are made of woven stainless steel mesh of 18 $\mu$m wires with 63 $\mu$m pitch, stretched over a readout anodic PCB of 10$\times$10 cm$^{2}$ active area segmented in 32$\times$32 square pads with 3.5 mm pitch. The MM gap is 128 $\mu$m, thanks to 500 $\mu$m diameter photo-resist pillars located at the center of each pad (Fig.\ref{fig:micromegasscheme}-D). 

\par
High Voltage (HV) bias is individually provided to each pad via a 470 M$\Omega$ protection resistor (Fig.\ref{fig:micromegasscheme}-B). External HV distribution cards hosting 128 such resistors each are mounted on the back of the anode PCB (Fig.\ref{fig:hybridminipadschematic}). Readout pads with the same geometry of the anodic pads (Fig.\ref{fig:micromegasscheme}-A), buried 70 $\mu$m inside the anode PCB provide the induced signal to the front-end electronics. The readout is based on the APV-25~\cite{APV25} and the Scalable Readout System (SRS) \cite{SRS} developed within the framework of the RD51 collaboration at CERN. The prototype readout system is fully modular and contained within the active area, allowing simple expansion of the design to cover larger areas.

\par
All MMs have been characterized using $^{55}$Fe X-Ray source and Ar:CO$_{2}$ 70:30 gas mixture and showed stable operation at effective gain above 10k.

\par
\begin{figure}
	\centering
	\includegraphics[width=0.7\linewidth]{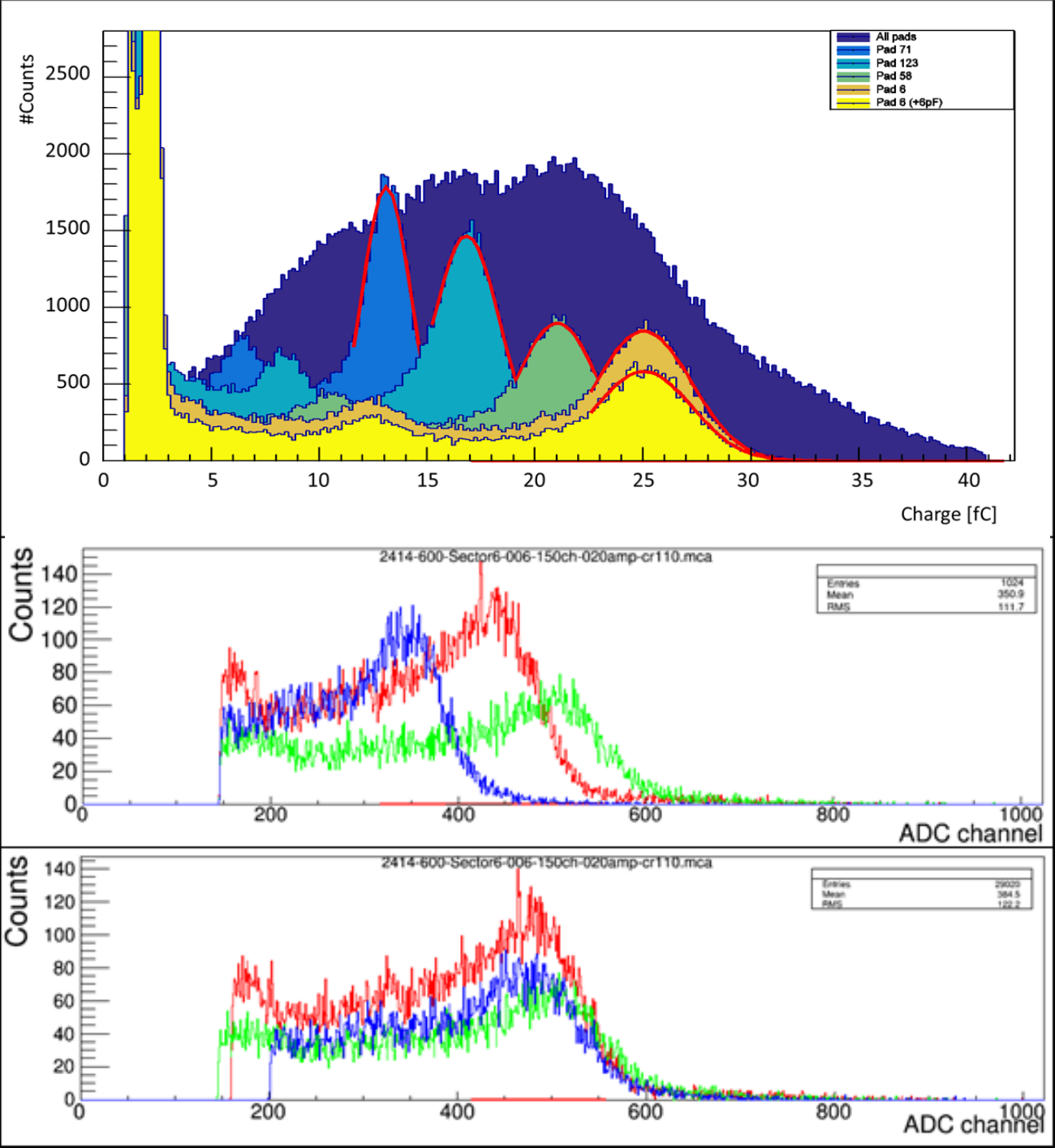}
	\caption{Top: $^{55}$Fe amplitude spectra collected by different pads using the MM multiplication stage only and the convolution of these distributions. 
	Middle: raw spectra from single pads using $^{55}$Fe X-Ray source. 
	Bottom: Amplitude spectra corrected according to the different measured parasitic
	capacitance affecting the pads.}
	\label{fig:mmcapacitanceproblem}
\end{figure}

During the characterization a distorted shape of amplitude spectra (while reading 128 pads together) was observed due to non uniform response, up to $\pm$20\%, among different pads. (Fig.\ref{fig:mmcapacitanceproblem}-Top). The source of non uniformity was identified: it is related to parasitic capacitance difference within pads; the same signal provided to the MM mesh induces different amplitude signals in the readout pads. 
After correcting the amplitudes for these measured 
parasitic capacitance differences, the amplitude spectra from the different pads provide similar amplitude peak positions (Fig.\ref*{fig:mmcapacitanceproblem}-Middle, Bottom).  

\section{Raven DAQ and Raven Decoder}

\par
\begin{figure}
	\centering
	\includegraphics[width=0.9\linewidth]{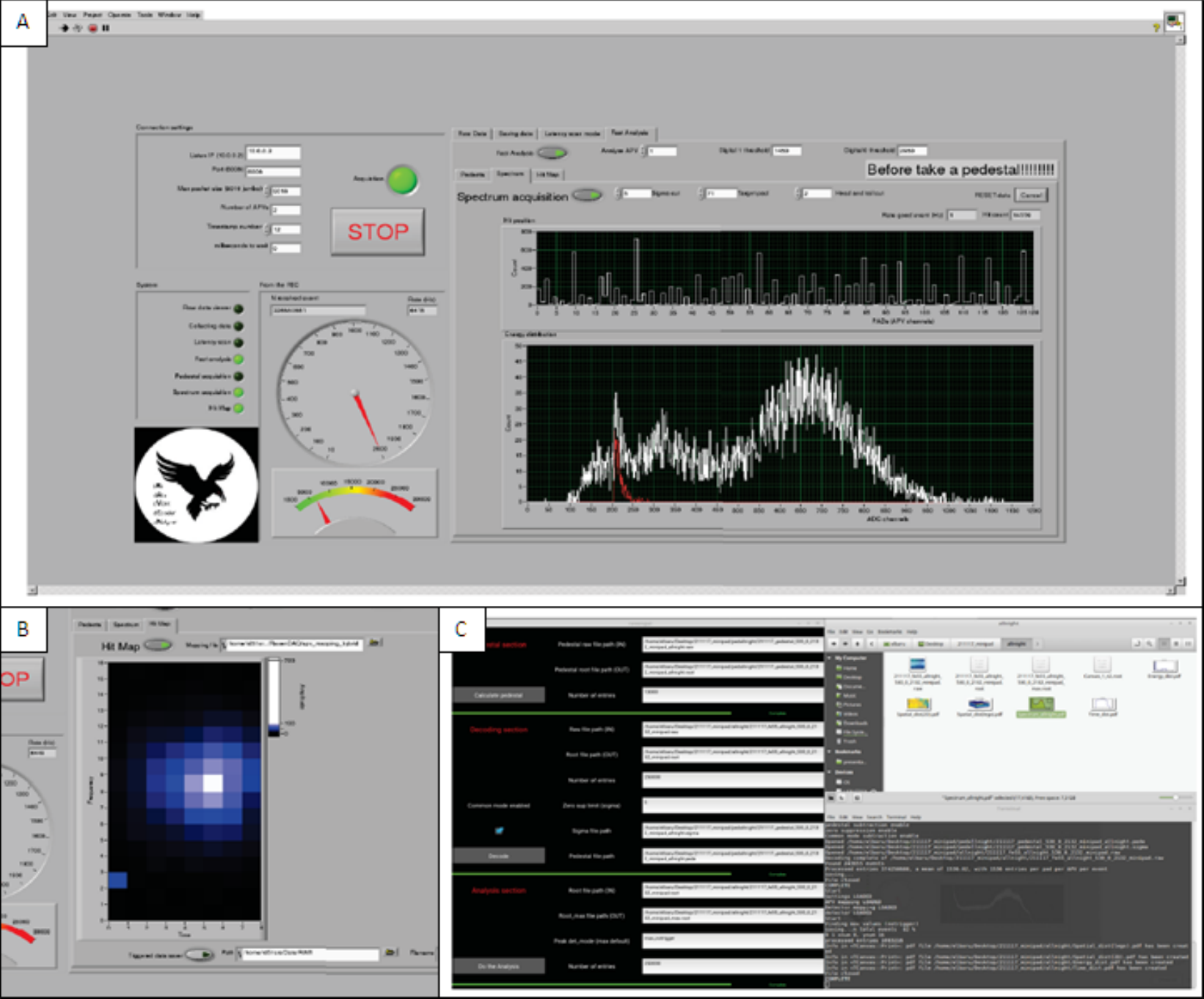}
	\caption{Raven DAQ GUI. A. The amplitude spectrum collected using a $^{55}$Fe X-Ray source to
	illuminate the prototype and reading the 128 channels of an APV chip is shown.
	B. Hit map of a 8$\times$16 pad matrix of a APV chip showing the hit map. C. The GUI showing the Raven decoder.}
	\label{fig:ravendaqanddecoder}
\end{figure}

To read and collect data from the APV-25 based SRS system used for the prototype, a simple, user friendly data acquisition system called Raven DAQ was developed using LabVIEW for laboratory and test beam  applications. 
To decode and analyze the saved data by Raven DAQ a C++ based graphical interface has been developed called Raven Decoder (Fig.\ref*{fig:ravendaqanddecoder}-C). Raven DAQ can perform zero suppression and allows high data acquisition rate: the performance depends on the number of APVs and the writing speed of PC hard disks (10 kHz for 1 APV with 7200 RPM SATA disks). 
The GUI (see Fig.\ref*{fig:ravendaqanddecoder}-A\&B) allows to perform on-line data analysis and essential tasks (e.g. Spectrum demonstrations, hit-map display, pedestal acquisition, latency scan etc.) in real time while saving the raw data to computer disks.

\section{Preliminary results} 


\par
The modular minipad hybrid PD prototype was assembled, tested and characterized at the INFN Trieste Photon Detector Laboratory. Using a large fused silica window and Picoquant PLD 4000B pulsed UV laser source and Ar:CH$_{4}$ 50:50 gas mixture, the gas mixture in use for COMPASS RICH hybrid PDs, a full characterization has been performed: stable operation above 50k effective gain was observed.

\par
\begin{figure}
	\centering
	\includegraphics[width=0.9\linewidth]{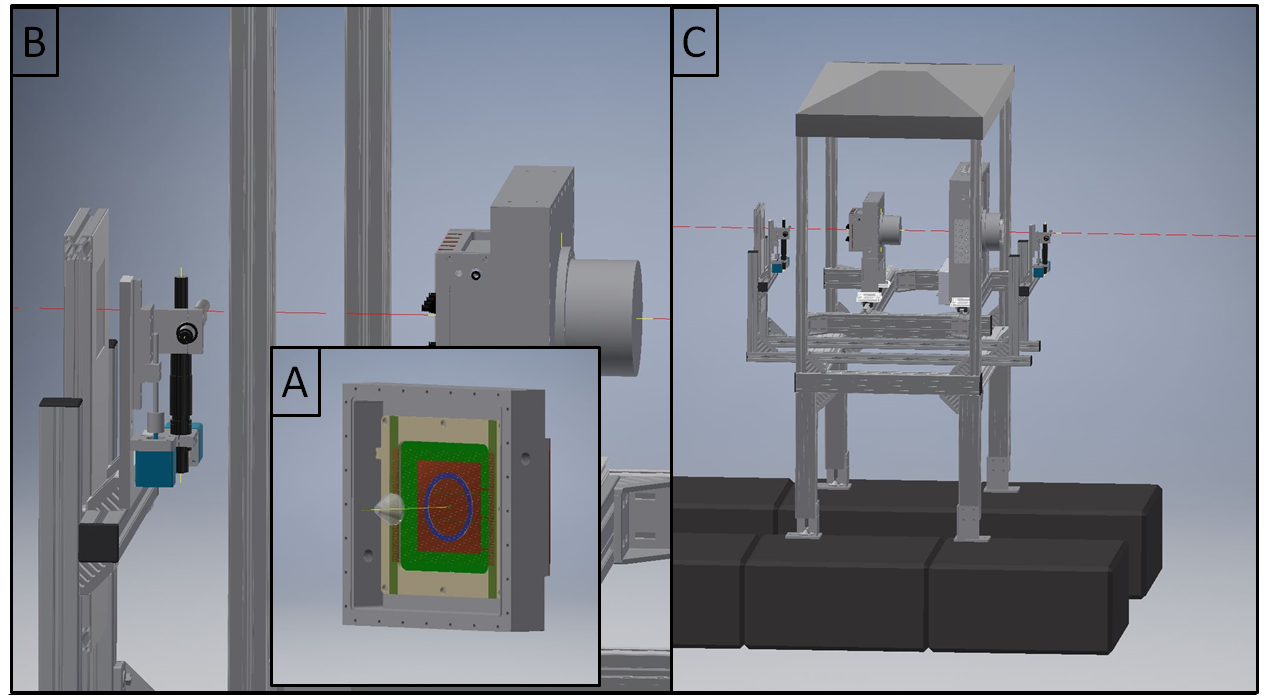}
	\caption{A: Scheme of the solid radiator and the detector; 
	B: The closeup sketch of the chamber with radiator housing in the test beam setup 
	with the trigger system; 
	C: The complete test beam setup.}
	\label{fig:testbeamsetup}
\end{figure}

A complete test beam exercise (Fig.\ref*{fig:testbeamsetup}) 
has been performed at the CERN H4 beam line using $\pi$ and $\mu$ beams focused on a truncated cone solid Cherenkov radiator aligned to the center of the pad plane. A remote controlled 
iris diaphragm  system  has been designed and implemented to control the amount of Cherenkov light. 

\begin{figure}
	\centering
	\includegraphics[width=0.9\linewidth]{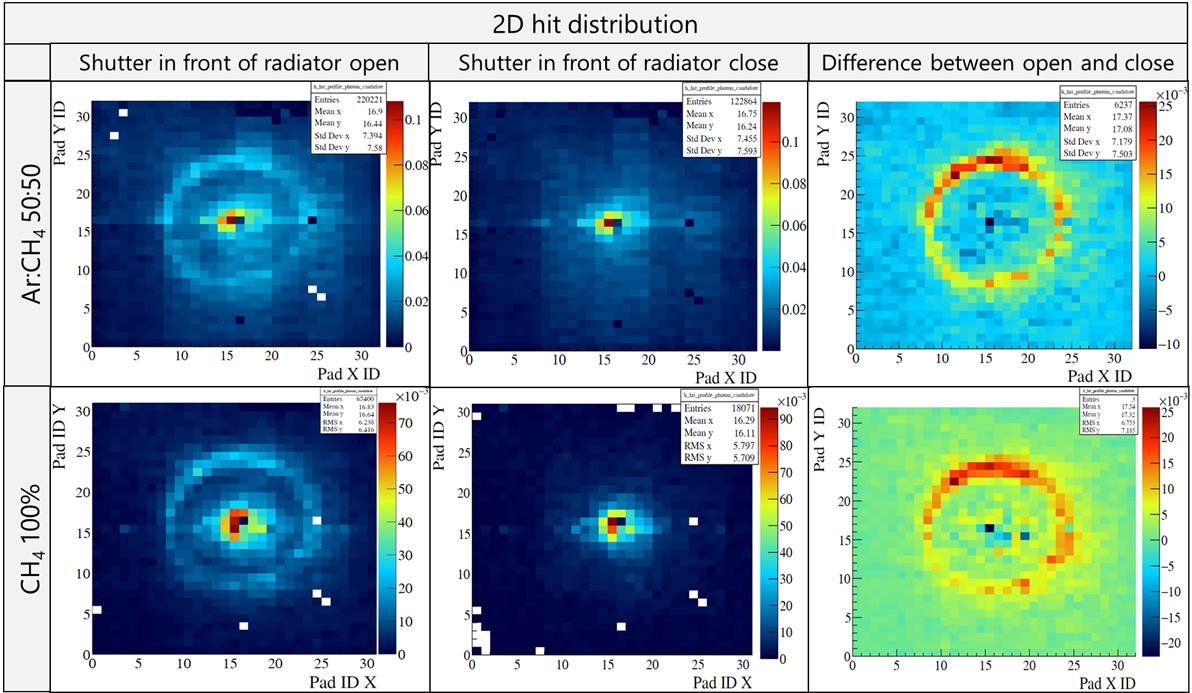}
	\caption{Observed Cherenkov rings in $Ar:CH_{4}$ 50:50 gas mixture and in pure $CH_{4}$ gas}
	\label{fig:CerenkovRings}
\end{figure}

With iris completely open clear Cherenkov rings has been observed in Ar:CH$_{4}$ 50:50 gas mixture and in pure methane gas (Fig.\ref*{fig:CerenkovRings}).

\section{Conclusion}

A modular hybrid minipad  PD of  100 $\times$ 100~$mm^{2}$ active area has been built and characterized in the laboratory and then in a test beam at CERN SPS. The observed non-uniform capacitance of the read out channel suggests a redesign of the anode PCB. The first prototype module was successfully tested in beam at CERN in October 2018. The promising first results show such a modular detector of single photon could provide high space resolution for single photon detection and can cover large areas.



\section{Acknowledgment}
\par
The activity is partially supported by the H2020 project AIDA2020 GA no. 654168. 
\par
It is supported in part by CERN/FIS-PAR/0007/2017 through COMPETE, FEDER and FCT (Lisbon). 
\par
The authors are member of the RD51 Collaboration: they are grateful to the Collaboration for the effective support and the precious encouragements.

\end{document}